\documentclass[10pt]{article}
\usepackage{amssymb}
\usepackage{amstext}
\newcommand*\de{\mathrm{d}}
 
\renewcommand*\epsilon{\varepsilon}
\renewcommand*\phi{\varphi}
\renewcommand*\theta{\vartheta}
\renewcommand*\P{{}^{\text{\tiny T}}\!\pi}
\newcommand*\PP{{}^{\text{\tiny TT}}\!\pi}
\newcommand*\HH{{}^{\text{\tiny TT} }\!h}
\begin{document}

\title{\bf \large Faddeev-Jackiw quantization of spin-2 field} 

\author{ M. Leclerc\footnote{mleclerc@phys.uoa.gr}\\ \small
  Section of Astrophysics and Astronomy, 
Department of Physics,\\ \small  University of Athens, Greece}  
 \date{\small December 20, 2006}
\maketitle
\begin{abstract}
 We apply the  Faddeev-Jackiw method to  the Hamiltonian analysis of 
the massless spin-two field. As expected, the reduced Hamiltonian 
contains only the traceless-transverse tensor, while some, but  not all 
of the non-propagating 
components are determined by the constraints of the theory. 
In particular, it is concluded that no gauge choice can be imposed on the 
fields such that only the propagating modes remain in the theory, meaning that 
for the spin-2 field 
there is no direct analogue to the Coulomb gauge of electromagnetism.  
Implications for General Relativity are discussed. 
\end{abstract}

\section{Introduction}

The Faddeev-Jackiw formalism \cite{faddeev} for constrained Hamiltonian 
systems is often the quicker alternative 
to the conventional Dirac method \cite{dirac}, because the introduction of 
unnecessary momentum variables is avoided and thus the number of constraints 
is reduced right from the outset. This holds in particular for first order 
theories, like Einstein-Cartan or Dirac theory. Here, we will take 
advantage of yet another feature  of this formalism, namely 
 the fact that the physical degrees of freedom are clearly 
identified without the need to impose a gauge fixation, 
and Poisson brackets are introduced only for those degrees of 
freedom. The idea behind the Dirac method was in a certain sense 
just the opposite, namely to 
treat all fields identically, at least initially, and to reduce the
Hamiltonian only  with respect to the second class constraints. First class 
constraints can be  directly imposed on the states without leading to 
inconsistencies. In this way, 
one can pass to the quantum theory without explicitely 
solving the constraints.  
While this is particularly convenient in theories where the constraints 
cannot be solved, as is the case, e.g., in canonical General Relativity 
\cite{dewitt}, it is nevertheless 
not very instructive as far as the identification of 
the physical modes, and thus of the particle contents 
is concerned. The theory, in the Dirac approach, 
can only be reduced to the physical modes by explicitely fixing the gauge. 

On the other hand, in the Faddeev-Jackiw approach, the constraints are 
explicitely solved (whenever possible) and the Hamiltonian is reduced to the 
physical degrees of freedom without choosing a gauge. 
This is very apparent in the case of the  electromagnetic field presented in 
\cite{faddeev}, where the Faddeev-Jackiw method leads straightforwardly to the 
Coulomb gauge Hamiltonian, which contains only the transverse spin-one field. 
This shows that  the Coulomb gauge is actually the 
 \textit{natural} choice, albeit not always the most convenient one. 
This holds  as long as we perform the Hamiltonian 
analysis with respect to a spacelike hypersurface. Other gauges appear, e.g., 
for a lightcone quantization.  

That only the physical degrees of freedom are quantized in the Coulomb gauge 
was known long before the advent of theoretical treatments of constrained 
Hamiltonian systems. However, in the case of the spin-two field, the 
situation is not quite as clear. What one would like to see is a gauge 
choice that leads to a wave equation for the traceless-transverse 
(3-dimensional) tensor, while the remaining 8 components of the symmetric 
(4-dimensional) tensor are either eliminated by a Gauss type law (as is 
$A_0$ in the Coulomb gauge) or put to zero by a convenient gauge choice 
(as is the longitudinal part of $A_{\mu}$ in the Coulomb gauge). While this 
should be possible on the base of a counting argument, the fact is that 
this has never been completely accomplished in practice. For instance, 
in the appendix of 
\cite{leclerc}, we have shown how in  the field 
equations of linearized General Relativity  one can 
make a gauge choice that allows for the elimination of 
 four components by a Gauss type law $\Delta \psi_{0i} = 
0$, $i = 0,1,2,3$, where $\psi_{ik}$ is the trace-reversed 
metric perturbation, while the remaining tensor is transverse,   
$\psi^{\mu\nu}_{\  \ ,\mu} = 0$, ($\mu,\nu = 1,2,3$),  and satisfies a 
wave equation. The remaining gauge freedom, however, is too restricted 
to eliminate yet another component of $\psi^{\mu\nu}$ in order to reduce it 
to the propagating  traceless-transverse spin-two field. This can only be 
done for explicit (e.g., plane) wave solutions. While this is sufficient to 
establish the nature of the field and to exclude, e.g., the existence of 
an additional scalar particle, it is nevertheless not really satisfying. 
In particular, one might wonder whether this is a result of a bad 
gauge choice, or whether it is truly impossible to exclude all 8
non-propagating components of $h_{ik}$. The Faddeev-Jackiw analysis should 
shed some light on the issue.   

\section{Faddeev-Jackiw reduction}

We start from the special 
relativistic Fierz-Pauli  Lagrangian, which is also the Lagrangian of 
General Relativity to second order in the metric perturbation $g_{ik} = 
\eta_{ik} + h_{ik}$, 
\begin{equation} \label{1}
L =  \frac{1}{4} h^{il}_{\  ,k} h_{il}^{\ \ ,k} - \frac{1}{2} 
h^{il}_{\  ,i} h_{kl}^{\ \ ,k} - \frac{1}{4} h^m_{\ m,l}h^{k\ ,l}_{\ k} 
  + \frac{1}{2} h^{km}_{\ \ ,k} h^l_{\ l,m}, 
\end{equation}
where latin subscripts  run from $0$ to $3$ and greek ones from $1$ to $3$. 
Our metric convention is $\eta_{ik}\! =\! diag(1,\!\!-1,\!\!-1,\!\!-1\!)\!$. 
Further, we will 
use the notation $\Box = \partial_i \partial^i$ and $\Delta = \partial_{\mu} 
\partial^{\mu}$. Whenever we write down a Lagrangian $L$ or a Hamiltonian $H$, 
an integration over three dimensional space is understood. 
Also, total 
time derivatives as well as two-dimensional surface terms are added or 
eliminated without explicitely renaming $L$ (or $H$). We choose as 
independent fields the components $h_{0\mu}, h_{00} $ and $h_{\mu\nu}$. 
 The corresponding momenta 
are found in the form 
\begin{eqnarray} \label{2} 
p^{00} = -\frac{1}{2} h^{\nu 0}_{\ \ ,\nu},\quad  p^{0\mu} = -h^{\nu\mu}_{\ \
  ,\nu} + \frac{1}{2} h^{,\mu} + \frac{1}{2} h^{00,\mu}, \\ \label{3}
p^{\mu\nu} =  \frac{1}{2} \dot h^{\mu\nu} - \frac{1}{2} \dot h \eta^{\mu\nu}
+ \frac{1}{2} h^{\lambda0}_{\ \ ,\lambda} \eta^{\mu\nu}, 
\end{eqnarray}
where $h = h^{\mu}_{\ \mu}$ is the trace of the 3-dimensional tensor, 
and the dot denotes  partial time derivatives. In the Dirac formalism, 
the relations (\ref{2}) would be considered as constraints. Here, they 
simply indicate that there are no momentum variables conjugated to 
$h_{0\mu}$ and $h_{00}$, and the symbols $p^{00}$ and $p^{0\mu}$ are only 
used as an abbreviation for the expressions on the right hand side. 
We introduce the Hamiltonian 
\begin{equation} \label{4} 
\hat H = p^{00} \dot h_{00} + p^{0\mu} \dot h_{0\mu} + p^{\mu\nu} 
\dot h_{\mu\nu} - L, 
\end{equation}
where the velocities $\dot h_{\mu\nu}$. 
are eliminated with the help of (\ref{3}). (Obviously, 
the remaining velocities, $\dot h_{0\mu}, \dot h_{00}$,  cancel out.)
The first order Lagrangian therefore is of the form 
\begin{equation} \label{5} 
L = p^{00} \dot h_{00} + p^{0\mu} \dot h_{0\mu} + p^{\mu\nu} \dot 
h_{\mu\nu} - \hat H(p_{\mu\nu}, h_{\mu\nu}, h_{0\mu}, h_{00}). 
\end{equation}
Using the explicit expressions for $p^{00}$ and $p^{0\mu}$, performing 
partial integrations and omitting surface terms and total time 
derivatives, we can alternatively write 
\begin{eqnarray} 
L &=& (p^{\mu\nu} - h^{0(\mu,\nu)} + \frac{1}{2} \eta^{\mu\nu} 
h^{0\lambda}_{\ \ ,\lambda})\ 
\dot 
h_{\mu\nu} - \hat H(p_{\mu\nu}, h_{\mu\nu}, h_{0\mu}, h_{00}) \nonumber 
\\ \label{6}
&=& \pi^{\mu\nu} \dot h_{\mu\nu} - \hat H( \pi_{\mu\nu}, h_{\mu\nu}, 
h_{0\lambda}, h_{00}), 
\end{eqnarray}
where we have defined the new momenta 
\begin{equation} \label{7}
\pi^{\mu\nu} = p^{\mu\nu} - h^{0(\mu,\nu)} + \frac{1}{2} \eta^{\mu\nu}
h^{0\lambda}_{\ \ ,\lambda}. 
\end{equation}
Explicitely, the Hamiltonian has the form 
\begin{eqnarray} \label{8}
 \hat H &=&  \pi^{\mu\nu} \pi_{\mu\nu} - \frac{1}{2} \pi^2 + \frac{1}{4} 
h_{,\mu} h^{,\mu} - \frac{1}{4} h^{\nu\lambda}_{\ \ ,\mu} 
h_{\nu\lambda}^{\ \ ,\mu}+ \frac{1}{2} 
h^{\mu\lambda}_{\ \ ,\mu} h_{\nu\lambda}^{\ \ ,\nu} 
- \frac{1}{2} h^{\mu\nu}_{\ \ ,\nu}h_{,\mu} \nonumber \\
&&   - h_{0 \mu} ( 2 \pi^{\mu\nu}_{\ \ ,\nu}) - 
h_{00}(-\frac{1}{2} h^{\mu\nu}_{\  \ ,\mu,\nu} 
+ \frac{1}{2} h^{,\mu}_{\ ,\mu}). 
\end{eqnarray}
The variables $h_{0\mu}$ and $h_{00}$ remain undetermined and play the 
role of Lagrange multipliers. Changing the notation from $h_{0\mu}$ 
to $\lambda_{\mu}$ and from $h_{00}$ to $\lambda_0$, we can write the 
Lagrangian in the form 
\begin{equation} \label{9} 
L = \pi^{\mu\nu} \dot h_{\mu\nu} - H(\pi^{\mu\nu}, h^{\mu\nu}) - 
\lambda_{\mu} \Phi^{\mu} - \lambda_0 \Phi^0, 
\end{equation}
where $H(\pi_{\mu\nu}, h_{\mu\nu})$ is the reduced Hamiltonian, given 
by the first line in (\ref{8}), and $\Phi^{\mu}$ and $\Phi^0$ are the 
constraints found in the second line of (\ref{8}), namely 
\begin{equation} \label{10}
\pi^{\mu\nu}_{\ \ ,\mu} = 0, \ \ h^{\mu\nu}_{\ \ ,\mu\nu} - \Delta h = 0. 
\end{equation}
In the Dirac approach, those constraints, together with the constraints 
(\ref{2}) (which are easily shown to be all first class, 
see also \cite{baaklini}) are imposed on the states of the quantum theory, 
and we could essentially stop the analysis at this stage (after checking 
that no tertiary constraints arise).  
This, however, 
does not really illuminate the situation as far as the physical degrees of 
freedom are concerned. On the other hand, the Faddeev-Jackiw procedure
consists in solving the constraints, thereby further 
reducing  the Hamiltonian. 

Let us  solve the constraints, at least formally. First, we see that 
the momentum $\pi^{\mu\nu}$ is transverse. In contrast to the vector case, 
however, the decomposition into a transverse and a longitudinal part of 
a symmetric tensor is not unique. 
It turns out though, that the explicit solution is not required for the 
moment. We simply 
\textit{solve} the constraint by writing  
$\pi^{\mu\nu} = \P^{\mu\nu}$, where $\P^{\mu\nu}$ is transverse.   
The second constraint can be solved for the trace $h$. Defining 
$\tilde h^{\mu\nu} = h^{\mu\nu} - \frac{1}{3} \eta^{\mu\nu} h$, which is 
the traceless part of $h^{\mu\nu}$, we find $\Delta h = (3/2) \tilde
h^{\mu\nu}_{\ \ ,\mu,\nu}$, or 
\begin{equation} \label{11}
h^{\mu\nu} = \tilde h^{\mu\nu} + \frac{1}{2} \eta^{\mu\nu}
\frac{\nabla_{\alpha}\nabla_{\beta}}{\Delta} \tilde h^{\alpha \beta},  
\end{equation}
where $\nabla_{\mu} \equiv \partial_{\mu}$ denotes partial differentiation.
The Lagrangian reduces to 
\begin{eqnarray} \label{12} 
 L &=& \P^{\mu\nu} (\dot {\tilde  h}_{\mu\nu}+ 
 \frac{1}{2} \eta^{\mu\nu}
\frac{\nabla_{\alpha}\nabla_{\beta}}{\Delta} \dot{ \tilde  h}^{\alpha \beta})
\nonumber \\ && 
- \left[  \P^{\mu\nu}\P_{\mu\nu} - \frac{1}{2} \P^2 - \frac{1}{4} \tilde 
h^{\nu\lambda}_{\ \ ,\mu} \tilde h_{\nu\lambda}^{\ \ ,\mu} + \frac{1}{2}
\tilde h^{\mu\lambda}_{\ \ ,\mu} \tilde h^{\alpha}_{\ \lambda,\alpha} 
+ \frac{1}{8} \tilde h^{\mu\nu}_{\ \ ,\mu,\nu}
 \frac{\nabla_{\alpha}\nabla_{\beta}}{\Delta} \tilde h^{\alpha\beta}
\right]. 
\end{eqnarray}
We have yet to bring the kinematical term into the canonical form. 
Obviously, $\P_{\mu\nu}$ cannot be conjugated to $\tilde h_{\mu\nu}$, 
because the number of independent components does not match. Fortunately, 
the reduction is straightforward. First, we define the traceless-transverse 
part of a symmetric tensor by 
\begin{eqnarray} \label{13} 
{}^{\text{\tiny TT}}\!A_{\mu\nu} &=& A_{\mu\nu} 
- \frac{\nabla_{\mu} \nabla^{\lambda}}{\Delta} 
A_{\lambda\nu}   
- \frac{\nabla_{\nu} \nabla^{\lambda}}{\Delta}  A_{\lambda\mu}   
+ \frac{1}{2} 
\eta_{\mu\nu} \frac{\nabla_{\alpha}\nabla_{\beta}}{\Delta} A^{\alpha\beta} 
\nonumber \\ &&
+ \frac{1}{2} \frac{\nabla_{\mu} \nabla_{\nu}}{\Delta} A 
+ \frac{1}{2}  \frac{\nabla^{\mu} \nabla^{\nu}}{\Delta} 
\frac{\nabla_{\alpha}\nabla_{\beta}}{\Delta} A^{\alpha\beta} 
- \frac{1}{2} \eta_{\mu\nu} A.  
\end{eqnarray}
This expression, in contrast to the merely transverse part, is unique, as is 
easily shown by considering a general linear combination of the terms 
occurring in (\ref{13}) and requiring 
${}^{\text{ \tiny TT}}\!A^{\mu\nu}_{\ \ ,\mu} = 0$ and 
${}^{\text {\tiny TT}}\!A = 0$. 
In particular, for $\P_{\mu\nu}$, which is already 
transverse, we have  
\begin{equation} \label{14}
\PP_{\mu\nu} = \P_{\mu\nu} + \frac{1}{2}
\frac{\nabla_{\mu}\nabla_{\nu}}{\Delta}\ \P - \frac{1}{2} \eta_{\mu\nu} \P. 
\end{equation}
It is not hard to show that, up to a  divergence, we have 
the following identity
\begin{equation} 
\PP_{\mu\nu} \, ( {\dot {\tilde h}}^{\mu\nu} + \frac{1}{2} \eta^{\mu\nu} 
\frac{\nabla_{\alpha} \nabla_{\beta}}{\Delta} 
{\dot {\tilde h}}^{\alpha\beta}) = 
\P_{\mu\nu} \, ( {\dot {\tilde h}}^{\mu\nu} + \frac{1}{2} \eta^{\mu\nu} 
\frac{\nabla_{\alpha} \nabla_{\beta}}{\Delta} 
{\dot {\tilde h}}^{\alpha\beta}). 
\end{equation}  
On the other hand, we also have $\PP_{\mu\nu} A^{\mu\nu} = 
\pi_{\mu\nu}  {}^{\text{\tiny TT}}\!\! A^{\mu\nu} 
= \PP_{\mu\nu}  {}^{\text{\tiny TT}}\!\! A^{\mu\nu} $ for any 
symmetric tensor $A^{\mu\nu}$ (up to divergence terms), 
and moreover, we have 
${}^{\text{\tiny TT}}\! \dot h^{\mu\nu} = {}^{\text{\tiny TT}}\!(
 {\dot {\tilde h}}^{\mu\nu} + \frac{1}{2} \eta^{\mu\nu} 
\frac{\nabla_{\alpha} \nabla_{\beta}}{\Delta} 
{\dot {\tilde h}}^{\alpha\beta})$. As a result, the kinetic 
term in (\ref{12}) is equivalent 
to $\PP_{\mu\nu} {}^{\text {\tiny TT}}\! \dot h^{\mu\nu}$, 
and the Lagrangian can be written as 
\begin{equation}\label{15}
L =  \PP_{\mu\nu} {}^{\text{\tiny TT}}\! \dot h^{\mu\nu} 
- H(\PP_{\mu\nu}, \HH_{\mu\nu}, \lambda), 
\end{equation}
where $\lambda$  denotes collectively the remaining parts of 
$\P_{\mu\nu}$ and $\tilde h_{\mu\nu}$. Those parts could lead to 
additional constraints and to a further reduction of the Hamiltonian. 
It turns out, however, that the Hamiltonian does not depend on the 
non-traceless-transverse components of $\tilde h_{\mu\nu}$ and $\P_{\mu\nu}$. 
Indeed, using (\ref{14}) as well as 
\begin{equation} \label{14b}
\HH_{\mu\nu} = \tilde h_{\mu\nu} - 
- \frac{\nabla_{\mu} \nabla^{\lambda}}{\Delta} 
\tilde h_{\lambda\nu}   
- \frac{\nabla_{\nu} \nabla^{\lambda}}{\Delta}  \tilde h_{\lambda\mu}   
+ \frac{1}{2} 
\eta_{\mu\nu} \frac{\nabla_{\alpha}\nabla_{\beta}}{\Delta} 
\tilde h^{\alpha\beta} 
+ \frac{1}{2}  \frac{\nabla^{\mu} \nabla^{\nu}}{\Delta} 
\frac{\nabla_{\alpha}\nabla_{\beta}}{\Delta} \tilde h^{\alpha\beta},  
\end{equation}   
where $\tilde h^{\mu\nu}$ is already traceless, we find,  
 up to surface terms, 
the simple expression 
\begin{equation}\label{16}
H =  \PP^{\mu\nu}\, \PP_{\mu\nu} - \frac{1}{4} \HH^{\mu\nu}_{\ \ ,\lambda}
\HH_{\mu\nu}^{\ \ ,\lambda}. 
\end{equation}
Note that one can also directly check that (\ref{8}) depends only 
on the traceless-transverse part of $h_{\mu\nu}$ (except for the 
expressions contained in the constraints), by checking the 
invariance of the $h_{\mu\nu}$-terms under $\delta h_{\mu\nu}
= \xi_{\mu,\nu}+\xi_{\nu,\mu}$. 
The Poisson brackets (at equal times) can be read off from (\ref{15})  and 
read 
\begin{equation} \label{17}
[\PP^{\mu\nu}(\vec x), \HH_{\alpha\beta}(\vec y)] 
= - \frac{1}{2}(\delta^\mu_\alpha 
\delta^{\nu}_{\beta} + \delta^{\nu}_{\alpha} \delta^{\mu}_{\beta}) 
\ \delta(\vec x - \vec y)
\end{equation}
The equations of motion derived from (\ref{16}) are 
\begin{equation} \label{18} 
 - {}^{\text{ \tiny TT}}\! \dot \pi^{\mu\nu} = [H, \PP^{\mu\nu}] 
=  \frac{1}{2}\Delta  \HH^{\mu\nu}, \quad
 -  {}^{\text{\tiny TT}}\! \dot h^{\mu\nu} = [H, \HH^{\mu\nu}]  
=- 2\, \PP^{\mu\nu}. 
\end{equation}
Putting this together leads to the expected equation for the propagating 
traceless-transverse tensor 
\begin{equation} \label{19}
\Box\, \HH^{\mu\nu}= 0. 
\end{equation}
As expected, the Faddeev-Jackiw method leads straightforwardly to the
identification of the propagating field modes. From (\ref{16}) and 
(\ref{17}), one can directly pass over to the quantum theory. Let us 
also note that our results are in accordance with those  obtained in 
\cite{baaklini} based on the Dirac method, after imposing convenient 
gauge conditions and evaluating the corresponding Dirac brackets. 
It should be noted, however, that in our treatment, no gauge conditions 
have been imposed. 

\section{Discussion}

While the above procedure was successful inasmuch the propagating 
components are concerned, a simple counting argument shows that we 
are missing something concerning the remaining field components. 
Indeed, according to (\ref{8}), $h_{00}$ and $h_{0\mu}$ remain 
undetermined. Obviously, this is a result of  the four 
gauge degrees of freedom of the theory (see below) and is a desired 
feature. Next, the 
two independent components of $\HH_{\mu\nu}$ satisfy the dynamical 
equation (\ref{19}). Thus, the remaining 4 components, which can be 
chosen as $h^{\lambda\mu}_{\ \ ,\mu}$ and $h$, should be 
completely determined by the constraints (\ref{10}). Those constraints, 
however, are not independent of each other and satisfy $\Phi^i_{,i} = 0$, 
which is a result of the linearized Bianchi identity, as we will see below. 
Thus, we have only three independent constraints for four fields, which 
means that one field component  remains undetermined. 

To show  this more explicitely, consider the field equations derived from 
the Lagrangian (\ref{1}),  
\begin{eqnarray} \label{21}
 0 \!\!&\!\!= G_{00} =\!\!&\!\!
 \frac{1}{2}(\Delta h - h^{\mu\nu}_{\ \ ,\mu,\nu}) \\
 0 \!\!&\!\!= G_{0\mu} =\!\!&\!\!
 \frac{1}{2}(- \Delta h_{0\mu} + \dot h^\nu_{\ \mu,\nu} 
+ h^{\nu0}_{\ \ ,\nu,\mu} - \dot h_{,\mu}) \label{22}\\ \label{23}
 0 \!\!&\!\!= G_{\mu\nu} =\!\!&\!\! - \frac{1}{2} \Box h_{\mu\nu} 
+ \frac{1}{2}(h^{\lambda}_{\ \nu,\lambda,\mu}
+ h^{\lambda}_{\ \mu,\lambda,\nu} - h_{,\nu,\mu} ) 
- \frac{1}{2}\eta_{\mu\nu}
(- \Box h + h^{\lambda\kappa}_{\ \ ,\lambda,\kappa}) \nonumber \\ 
&& \!\!
+ \frac{1}{2}(\dot h_{0\mu,\nu} + \dot h_{0\nu,\mu} - h_{00,\nu,\mu})
- \frac{1}{2} \eta_{\mu\nu}(-\Delta h_{00} 
+ 2 \dot h^{0\lambda}_{\ \ ,\lambda}),  
\end{eqnarray}
where $G_{ik}$ corresponds to  the linearized Einstein tensor. Before we 
continue, let us recall that the above equations are invariant under 
the gauge transformation 
\begin{equation} \label{24}
\delta h_{ik} = \xi_{i,k} + \xi_{k,i}, 
\end{equation}
which can be interpreted as the residual of the diffeomorphism invariance of 
General Relativity after linearization. In the Dirac approach, 
where Poisson brackets are defined for the complete set of fields, 
e.g.,   $[\pi_{\mu\nu}(x), h^{\alpha\beta}(y)] = -
\delta^{(\alpha}_\nu\delta^{\beta)}_\mu \delta(x-y)$, this symmetry is 
reflected in the occurrence of the constraints (\ref{10}). For 
instance, we have 
$[\int \xi_{\mu}2 \pi^{\mu\nu}_{\ \ ,\nu} \de^3 x,\,  h_{\alpha\beta}]= 
\xi_{\alpha,\beta} + \xi_{\beta,\alpha} = \delta h_{\alpha\beta}$, 
while $[\int \xi^0(-h^{\mu\nu}_{\ \ ,\mu,\nu}+ \Delta h)\de^3 x,\,
\pi_{\alpha\beta}] = -\xi^0_{\ ,\alpha,\beta} + \eta_{\alpha\beta} 
\Delta \xi^0 $, which is (on-shell) equal to  
$\delta \pi_{\alpha\beta}$ derived from (\ref{24}) using the 
explicit expression for  $\pi_{\alpha\beta}$ in terms of the fields, 
equations (\ref{3}) and (\ref{7}). 

Let us return to the field equations. First, from 
 the explicit expression for $\pi_{\alpha\beta}$,  we recognize in 
(\ref{21}) and (\ref{22}) the constraints (\ref{10}). Those equations do 
not contain second time derivatives. Next, with the projection 
given in  (\ref{13}), we take the traceless-transverse part of (\ref{23}) 
and find 
\begin{equation} \label{25} 
0 =   
{}^{\text{\tiny TT}}\! G_{\mu\nu}  = - \frac{1}{2} \Box\, \HH_{\mu\nu}, 
\end{equation}
which is the equivalent to our Hamiltonian equations, see  (\ref{19}). 
This leaves us with the 
longitudinal and trace components of $G_{\mu\nu}$. However, as is well known, 
the field equations satisfy the identity $G^{ik}_{\ \ ,k} = 0$ (linearized, 
contracted Bianchi identity) and therefore, we find $G^{\mu\nu}_{\ \ ,\nu} 
= - \dot G^{\mu0}$, which is zero as a result of (\ref{22}). Thus, 
$G^{\mu\nu}$ is already transverse as a consequence of the constraint
equations. Finally, we take the trace $G = G^{\mu}_{\ \mu}$ and find 
\begin{equation} \label{26}
G =0=  G_{00} + \ddot h + \Delta h_{00} - 2 h^{0\lambda}_{\ \ ,0,\lambda},
\end{equation}
which gives us one more equation 
\begin{equation} \label{27}
\ddot h + \Delta h_{00} - 2 h^{0\lambda}_{\ \ ,0,\lambda} = 0. 
\end{equation}
This equation did not appear in the previous section. Being the only 
equation that involves $h_{00}$, it is certainly independent of 
the remaining ones. Moreover, 
it is immediately clear from (\ref{26}) that it is gauge invariant, 
since both $G$ and $G_{00}$ are gauge invariant. It looks like a  dynamical
equation, since it contains second order time derivatives. 

To make thinks very explicit, let us choose the gauge 
$h_{00} = h_{0\mu} = 0$. It is clear from (\ref{24}) that the gauge 
is now quite  fixed (although transformations $\xi^{\mu}$ with $\dot \xi^{\mu} 
= 0$ are still allowed, but those leave $\dot h$ invariant).  
We then have the constraints $G^{0\mu} = G^{00} = 0$, 
which represent 
four relations between the four non-traceless-transverse components of 
$h_{\mu\nu}$ (which are not independent though, since $G^{0i}_{\ \ ,i} = 0$). 
Further, we have ${}^{\text{\tiny TT}}\! G_{\mu\nu}  = 0$, as well as 
a remaining equation $\ddot h = 0$, which was missing in our Hamiltonian 
analysis. As mentioned above, this equation cannot be gauged away. 
The fact that the constraints are not independent is obviously 
the reason why four constraints and four gauge choices cannot determine 
completely the 8 non-propagating components of $h_{ik}$, as long as we 
miss the equation $\ddot h = 0$.  

Note that in the Dirac approach, where the constraints are not eliminated
but rather imposed on the states of the quantum theory, we get the 
missing relation as dynamical equation of motion from $[H, \pi] = 
-\Delta h_{00} - G_{00}$, with the Hamiltonian $H(\pi_{\mu\nu}, h_{\mu\nu})$ 
from (\ref{9}). So, neither the Lagrangian field equations, which contain 
$\ddot h$, nor the Dirac analysis, which leads to an  additional dynamical 
equation, actually reveal  the fact that $h$ is  a non-propagating field. 
In other words, it is not obvious in those approaches that our Lagrangian 
describes only a spin-two field, and not, e.g., an additional scalar
particle. 

On the other hand, the Faddeev-Jackiw formalism shows  straightforwardly 
the true particle contents of the theory, but the price we have to pay 
is that one of the remaining field equations is mysteriously lost. 
One might think that we did something wrong during our calculations. 
This is not the case, however. The fact is that it is 
not possible to get the missing equation  in the Faddeev-Jackiw 
approach for the following reason. The formalism consists in reducing 
the Hamiltonian to a form which contains only the truly dynamical 
fields, and only for those fields Poisson brackets, and thus, ultimately, 
commutators, are introduced. In our case, this is undoubtlessly the 
traceless-transverse tensor $\HH_{\mu\nu}$ describing the spin-two particle.
Thus, the remaining components of $h_{ik}$ cannot be determined by dynamical 
equations of the form $[H,A] = - \dot A$. The only way we can determine  
 those components is therefore by the constraints of the theory. It is 
clear, however, that an equation containing second time derivatives (e.g., 
$ \ddot h$) cannot appear  
in the form of a constraint, because it necessarily contains 
velocities (e.g., $\dot \pi$). Thus, there is no way, in the 
framework of the Faddeev-Jackiw formalism,  to obtain 
equation (\ref{27}), which is seemingly dynamical, but nevertheless 
relates only  non-propagating fields. 

The fact that the Faddeev-Jackiw method sometimes leads to a loss of 
information  has been demonstrated by Garc\'ia and Pons 
in \cite{pons}. The specific case where this information represents a 
dynamical equation (and not a constraint) is a subcase of what is 
referred to as a type 2 problem in \cite{pons}. The authors  
 relate the failure of the Faddeev-Jackiw method  to the 
occurrence of so-called ineffective constraints. In our case, the situation 
seems different though. We do not go into the 
mathematical details of the analysis. Instead, we wish to point out an 
interesting conjecture put forth in \cite{pons}. Namely, the authors 
believe that, when the Dirac conjecture fails, then the Faddeev-Jackiw 
method must fail too. 
Thus, if the Dirac conjecture fails in the spin-two 
theory, then this not only provides an additional support for the 
Garc\'ia-Pons conjecture, but moreover, it shows that there are actually 
{\it natural} counterexamples to Dirac's conjecture, as opposed to the 
pathological theories whose only purpose is to  make the conjecture fail. 

It is clear that the loss of information was the result of omitting 
the constraints in the Hamiltonian (\ref{8}). By doing though, the 
remaining Hamiltonian depends only on the traceless-transverse fields, 
and we loose the equation $[H,\pi] = - \Delta h_{00}$ (on the constraint 
surface). This shows in particular that it is not allowed to replace 
$h_{00}$ and $h_{0\mu}$ in (\ref{8}) by arbitrary multipliers, i.e., 
to couple the secondary (in the Dirac scheme) constraints by 
Lagrange multipliers. Dirac's conjecture thus seems to be violated, 
but in a somewhat different sense.  
For a detailed investigation, one will have to check that 
the generator formed as a linear combination (with independent multipliers)
of the eight constraints (\ref{2}) and (\ref{10}) leaves invariant the 
physical quantities. Physical quantities are those that are invariant 
under the gauge transformation (\ref{23}), in particular the tensor 
components $G_{00}, G_{0\mu}$ and $G_{\mu\nu}$. For a detailed discussion 
of Dirac's conjecture in the context of gauge theories, 
we refer to  \cite{costa}. 
We find indeed that 
 equation (\ref{26}) (expressed in terms of $\pi$) is not
invariant under the action of this generator. (Note that  the remaining 
equations are trivially invariant: The constraint equations are first class, 
and thus commutate with each term of the generator separately, while 
the propagating equation involves only traceless-transverse components, 
which are not contained in the constraints, and thus neither in the 
generator.) The strange thing, however, 
is that (\ref{26}) 
it is not even invariant under the action of the generator formed 
merely with the primary constraints (\ref{2}). The reason is that 
the only constraint that involves $p_{0\mu}$, and thus the only constraint 
that can induce a transformation on $h_{0\mu}$, does not generate the 
transformation dictated by (\ref{26}). On the other hand, since the 
propagating fields $\PP_{\mu\nu}$ and $\HH_{\mu\nu}$ are indeed invariant 
under the full generator, in a certain sense, Dirac's conjecture holds 
true. In contrast to electrodynamics, however, where both $\pi^{\mu}$ and 
$F_{\mu\nu}$ are invariant \cite{costa}, and thus the complete 
tensor $F_{\mu\nu}$, 
in our case, there are quantities (albeit non-propagating ones) that are not 
invariant under the generator formed from the constraints although they 
 are invariant under (\ref{26}) when 
expressed in terms of configuration space variables. Such quantities 
cannot be discarded as unphysical, and we must conclude that, in this 
somewhat modified sense, the Dirac conjecture is violated. As outlined 
above, this is the reason why the replacement of $h_{00}$ and $h_{0\mu}$ by 
Lagrange multipliers (or simply the omission of the constraints) 
in the Hamiltonian (\ref{8}) is not permitted and leads to a loss 
of information. 
 
Finally, we wish to stress once again that, 
although the Faddeev-Jackiw approach 
has {\it failed} in the sense that some (seemingly) dynamical information 
has been  lost, there is no problem concerning the  propagating modes. 
The formalism works perfectly well in order to determine the physical 
modes and thus is perfectly suited for the transition to the quantum 
theory. In \cite{pons}, it has been shown that either constraints or 
{\it dynamical} equations might get lost in the Faddeev-Jackiw analysis.  
 The spin-two case presented here 
tells us that those {\it dynamical} equations are actually not quite as 
dynamical as they seem, since they 
involve only non-propagating field components.

We conclude  that 
the answer to the question raised in the 
introduction section is  negative.  It is 
{\bf not} possible to choose the gauge in a way that 
for some tensor $\psi_{ik}$ (which need not be equal to  $h_{ik}$ nor 
 to the trace-reversed tensor $h_{ik}- \frac{1}{2} \eta_{ik} h^m_{\ m}$, 
but could be an arbitrary  combination of components of $h_{ik}$), we 
have $\Delta \psi_{00} = \Delta \psi_{0\mu}  = 0 $ (i.e., a Gauss type 
law for the free field case), and such that in addition, 
$\psi_{\mu\nu}$ is traceless-transverse and satisfies a wave equation. 
Independently of the gauge we impose, eliminating thereby 4 components, 
the resulting field equations will still contain 3 (and not 2) accelerations. 
Thus, apart from the propagating field modes, there will  remain one 
more equation that cannot be eliminated by a Gauss type law. In the 
Hamiltonian formulation, this means that 
 3 (and not 2) pairs of Hamiltonian field equations are required 
in order to determine the configuration completely. 
The naive counting argument (10 components, 4 Gauss type eliminations and 
4 gauge degrees of freedom, 2 propagating components) does not 
work in the strict sense. In other words, there is no Coulomb type gauge 
for the spin-two field. One should  not confuse the situation with 
the corresponding problems arising in non-abelian gauge theories. In our 
 case, e.g., it is impossible to impose $h^{\mu\nu}_{\ \ ,\nu}  = 0$ 
together with $h = 0$ directly on the field in order to eliminate 
unphysical components. On the contrary, in the case of gauge theory, the 
condition $A^{\alpha\mu}_{\ \  ,\mu}$ ($\alpha$ is the internal group index) 
can  be imposed, but it does not fix the gauge completely 
(so-called Gribov ambiguity). This is more of a technical matter, meaning 
that the transversality condition of the free spin-one field cannot be 
 generalized to the self-interacting case without modifications. 
The principal situation,  namely that we have one constraint and two 
propagating modes (for each value of the $\alpha$) remains unaffected 
by this. 

Finally, one might come up with the idea to change the Lagrangian, in order 
to avoid the above problems. A class of alternative spin-two theories has 
been considered in \cite{blas}. 
The simplest modification of the theory  would consist in 
starting with a traceless field $h_{ik}$ right form the start. This 
corresponds to the linear approximation of the so-called unimodular 
General Relativity, see \cite{blas}. 
It  still leads to a relativistic spin-two theory, but with a number of  
field equations that is reduced by one. Hence, there is the hope that 
the Lagrangian field equations are now equivalent to the set of 
propagating and  constraint equations obtained in the Faddeev-Jackiw
reduction. This is not the case, though. Indeed, one finds that the number 
of constraints is reduced by one too, and there will again be one equation 
that gets lost during the reduction. The reason is that the gauge 
transformations $\xi^i$ now have to satisfy $\xi^i_{\ ,i}  = 0$ and thus, 
there are essentially only three gauge degrees of freedom, and hence three 
constraints. 

\section{Outlook: General Relativity} 

Since it is difficult to couple the spin-two field consistently to 
other fields, in particular to gravity, the main  motivation for its study 
is the fact that it is also  the first order approximation to General 
Relativity. It is therefore of interest to relate the above 
discussion to the case of the generally covariant theory. Obviously, 
from a perturbative point of view, there is not much to be expected. 
Treating gravity as a spin-two field on a flat background will 
reproduce to lowest order the above results, and only higher 
order modifications will result from the selfcouplings. The more interesting 
approach is certainly the canonical theory \cite{dewitt}.
Our question is thus the following. Suppose we were able, at least in 
principle, to solve the constraints of General Relativity, and 
to perform  the Faddeev-Jackiw reduction of the Hamiltonian, eliminating 
all non-propagating field modes. Would we loose again a dynamical 
field equation, as in the case of the spin-two field? In other words, 
is the field completely determined by the constraints and the propagating 
modes, or is there an additional dynamical equation? 

Before we try to gain some insight into that matter, let us 
consider a simpler example that permits us to see what would be 
the consequences of one or the other answer to the above question. 
Covariant theories are characterized by a vanishing Hamiltonian. Their 
mechanical counterpart are theories invariant under reparameterization. 
Therefore, let us consider the following special-relativistic point
particle theory
\begin{equation}\label{28b}
L = - m \sqrt{\dot x^{i}\dot x_{i}},  
\end{equation}
where the dot means derivation with respect to an unspecified curve 
parameter $\tau$, and the action is given by $S = \int L \de \tau$. 
The momenta satisfy $p_ip^i = m^2$, which can be solved 
for $p_0$ such that the  first order Lagrangian takes the  form  
\begin{equation} \label{29}
L = p_{\mu} \dot x^{\mu} + \sqrt{m^2 - p_{\mu}p^{\mu}}\  \dot x^0 
 - H(p_{\mu}, x^{\mu}, x^0). 
\end{equation}
To bring this into canonical form, we define 
\begin{equation}\label{30}
X^{\mu} = x^{\mu} - \frac{p^{\mu} x^0}{\sqrt{m^2 - p_{\mu}p^{\mu}}}, 
\end{equation}
which leads, up to a total $\tau$-derivative,  to 
\begin{equation} \label{31}
L = p_{\mu} \dot X^{\mu} - H(p^{\mu}, X^{\mu}, x^0).  
\end{equation}
Next, we have to eliminate the non-canonical variable $x^0$ by its 
equation of motion, $\partial H/\partial x^0 = 0$. 
However, the Hamiltonian is zero, and therefore, we are left with the 
equations of motion  
 $\dot X^{\mu} = \dot p_{\mu} =  0$. 
In particlar, the first equation can be trivially integrated, and 
from (\ref{30}), we find 
$x^{\mu} = \frac{p^{\mu} x^0}{\sqrt{m^2 - p_{\mu}p^{\mu}}} + a^{\mu}$, 
with a constant $a^{\mu}$. Since $p_{\mu}$ is constant too, this 
simply means $x^{\mu} = v^{\mu} x^0 + a^{\mu}$ with constant $v^{\mu}$, 
corresponding to free particle motion.  

What can we learn from this  example in relation to General Relativity? 
First of all, let us point out that, despite the fact that $H = 0$, 
there is nothing static about the above system. For instance, 
neither $\de x^{\mu}/\de \tau$, nor $\de x^{\mu}/\de x^0$ 
need to be zero. On the 
other hand, the canonical variables $ X^{\mu}, p^{\mu}$ are indeed constant. 
So, let us suppose we can devise a method to solve the constraints 
of General Relativity, at least formally, and to reduce the Hamiltonian 
to the physical variables (fields and momenta), which we shall denote by 
${}^{\text{\tiny TT}}\! g_{\mu\nu}$ and ${}^{\text{\tiny TT}}\! \pi_{\mu\nu}$, 
without having in mind that they are in some sense  transverse, 
nor traceless (or rather unimodular), 
such that we have 
\begin{equation} \label{32}
L = {}^{\text{\tiny TT}}\! \pi_{\mu\nu}  {}^{\text{\tiny TT}}\! \dot g_{\mu\nu}
- H({}^{\text{\tiny TT}}\! g_{\mu\nu},{}^{\text{\tiny TT}}\! \pi_{\mu\nu} ). 
\end{equation}
On the other hand, we know \cite{dewitt} that in General Relativity, 
we have four constraints, which correspond again to the Einstein equations 
$G^{0i} = 0$, and that, on the constraint surface, the Hamiltonian vanishes. 
Thus, we have $H = 0$, and consequently, the dynamical equations read 
${}^{\text{\tiny TT}}\! \dot g_{\mu\nu} = 0$ and 
${}^{\text{\tiny TT}}\! \dot \pi_{\mu\nu} = 0$. This apparent static 
nature of the theory has been studied intensively in literature, see, e.g., 
\cite{komar} and \cite{dewitt}. 
In spite of this, we will nevertheless continue to refer to 
those modes as {\it propagating}, whatever this means without reference to 
a background spacetime. Since they are the only modes that are subject to 
quantization (assuming that the standard procedures are applicable 
to General Relativity), we must conclude that they describe the spin-two 
particle. (Meaning that they contain the propagating modes $\HH_{\mu\nu}$ 
previously found in the special-relativistic theory. Without reference to 
a background space, the notion of a particle does not make sense anyway.)

Thus, if everything works out as planned, our system will be determined 
by the dynamical equations 
${}^{\text{\tiny TT}}\! \dot g_{\mu\nu} = 0$ and 
${}^{\text{\tiny TT}}\! \dot \pi_{\mu\nu} = 0$, as well as by the four 
constraints $G^{0i} = 0$. But what, if this is not the case? What,  if 
we have lost a dynamical equation during the Faddeev-Jackiw reduction?  
In that case, we would have an additional equation, say $\dot \pi = 
f(g_{ik}, \pi_{ik})$, and things would not be so static after all. 
In particular, this would mean that the standard approach is not entirely 
correct. Namely, it is traditionally assumed, since the dynamical 
equations are  trivial as a result of $H  = 0$,  that the complete 
information is contained in the four constraints $G^{0i}$, see \cite{dewitt} 
for details. But this might not be the case, as the example of the 
spin-two particle shows. As we have outlined in the previous section, 
the critical point is the replacement of $h_{00}$ and $h_{0\mu}$ 
that appear in front of the constraints,  
by arbitrary multipliers, or, alternatively, the omission of the 
constraints. In the same way, however, one deals with $g_{00}$ 
and $g_{0\mu}$ (or the lapse and shift functions $N,N_{\mu}$,  respectively, 
see \cite{dewitt}) in the canonical approach to  gravity. It could 
happen that this is not allowed  and leads to a loss of information. 

It is not easy to estimate whether we have indeed such a situation, 
but on the basis of a counting argument, it seems not improbable. 
First of all, there are again four undetermined components $g_{00}$ and 
$g_{0\mu}$, which is a result of the general coordinate invariance. 
Further, we have four constraints $G^{0i}$, related via the 
Bianchi identity $G^{0i}_{\ \ ;i} = 0$, where the semicolon denotes 
covariant derivation in four dimensions. On the constraint surface, 
this leads to $\Gamma^0_{\mu\nu} G^{\mu\nu} =  0$, where $\Gamma^i_{kl}$ 
is the Christoffel connection formed from the four dimensional metric.  
The remaining part, $G^{\mu\nu}$, contains the dynamical components. 
On the constraint surface, we find from $G^{\mu i}_{\ \ ;i}     = 0$ 
the relation $G^{\mu\nu}_{\ \ ,\nu} + \Gamma^{\mu}_{\lambda \nu} G^{\lambda\nu}
+ \Gamma^i_{\nu i} G^{\mu\nu} = 0$. We see that thinks are a lot more 
interrelated than  in the spin-two case, but the basic situation is the same. 
The relations  $G^{\mu\nu} = 0$ contain 3 independent equations, 
while from the four constraints $G^{0i} = 0$, only 3 are independent. 
As a result, there must be three equations containing second time derivatives 
(or, in the Hamiltonian formulation, three  canonically conjugated pairs of 
equations). 

That the Einstein equations contain three independent 
accelerations can be checked explicitely. Consider, e.g., the metric 
$\de s^2 = \de t^2 - g_{\mu\nu} \de x^{\mu} \de x^{\nu}$. Since $g_{\mu0}$ and 
$g_{00}$ are already fixed, only static, spatial  coordinate 
transformations are still allowed. Let us choose $g_{\mu\nu} = S(x) 
\tilde g_{\mu\nu}$, with 
$\tilde g_{xx} = 1-f(x)$, 
$\tilde g_{yy} = 1 + f(x)$, $\tilde g_{xy} =  g(x)$,   $\tilde 
g_{zz} = 1/(1-f(x)^2 - g(x)^2)$, $\tilde g_{xz} = \tilde g_{yz} = 0$. 
Note that $\tilde g_{\mu\nu}$ is unimodular and  $S(x)$ is essentially a
cosmological function. One can now directly check that $G_{00}$ and 
$G_{0\mu}$ do not involve second time derivatives. On the other hand, 
$G_{\mu\nu}$ contains $\ddot S(x), \ddot f(x)$ and $\ddot g(x)$. 
Since the coordinate system is already fixed with respect to time, this 
means that we have indeed three physically relevant functions that enter 
the field equations by their accelerations. (Note that it cannot be 
more than three neither, since we can still diagonalize $g_{\mu\nu}$.) 
We have chosen the particular form of $g_{\mu\nu}$ because $S$ is 
related to $\det g_{\mu\nu}$, and thus to the trace  $h$ in the 
linearized theory. Therefore, if the situation is similar to the linearized 
theory, we would have to conclude 
 that $S(x)$ is {\bf not} a propagating field, 
and in particular, is not subject to quantization. Such a result is 
certainly of relevance to quantum cosmology.

The question is therefore reduced to whether 
 after the complete reduction of the Hamiltonian, we still have 
three propagating modes contained in ${}^{\text{\tiny TT}}\! g_{\mu\nu}$, 
or whether we have only two, as seems to be suggested by the perturbative 
approach. The second case would result if we loose one equation 
during the reduction procedure (e.g., the equation containing $\ddot S$). 
This might appear to be an issue  without contents, since 
the so-called propagating modes satisfy 
${}^{\text{\tiny TT}}\! \dot g_{\mu\nu} = 0$ anyway and are no more and 
no less {\it propagating} than the remaining components. It is, however, 
of fundamental relevance,  since after all, it should be of 
importance to know  how many degrees of freedom are actually quantized. 

Without giving a definite answer to the above question, let us just 
say that the situation looks problematic in either way. In the second  
case, where we have only two propagating modes, we have a situation 
similar to the spin-two theory. That is, the two dynamical equations 
(which are trivial in the case of General Relativity), together with 
the four (not independent) constraints, do not describe the system 
completely. As we have outlined above, this means that the standard 
approach is not completely correct, i.e., the constraints do not 
completely determine the dynamics of the quantum system, 
in contrast to what  is claimed, e.g., in \cite{dewitt}. 
Rather, there is one more dynamical equation,  involving, e.g., the 
time derivative   of the momentum conjugated to $S$. 
In particular, this could mean that not everything needs to be 
 static after all. (For the precise meaning of the word {\it static} in 
this context, we refer to the interpretation given in  \cite{dewitt}.) 

In the other case, we would have three propagating modes, that is, 
three independent degrees of freedom 
are to be quantized. In that case, the conventional 
treatment would be valid, and, since the dynamical equations are trivial, 
the theory could be fully described by the constraints alone.  
It does not make sense to ask what kind of particles are described 
by such a theory, since one would need a reference background to answer that. 
The true question is, what happens to the third degree of freedom in 
the perturbative approach? How can we end up, to lowest  order, with a 
pure spin-two theory, if we start from a theory with three 
independent physical degrees of freedom?

Having at least a little amount of  faith in 
the perturbative approach, we would rather exclude that last  possibility and 
tend to believe that General Relativity is a spin-two theory. A further 
hint for the similarity to the linear theory comes from the fact that 
one can directly construct examples which demonstrate that the theory 
is not completely described by the constraints and the propagating modes 
alone. Although we do not  have identified the propagating modes exactly, 
we know (see \cite{dewitt}) that they are contained in the three-dimensional 
tensors $g_{\mu\nu}$ and $\pi^{\mu\nu}$. It is now easy to check, for 
instance, that the solution $g_{\mu\nu} = \eta_{\mu\nu}$, 
$\pi^{\mu\nu}  = 0$,  and 
$g_{0\mu} = 0$, $g_{00} = \exp[-r]$ satisfies the constraints and must 
also satisfy the (trivial) dynamical equations for the propagating parts 
contained in $\pi^{\mu\nu}$ and $g_{\mu\nu}$ (since $\dot g_{\mu\nu} = 
\dot \pi^{\mu\nu}=0$ anyway). Nevertheless, the above is not a solution 
of the Einstein equations. Thus, we are missing something.

\section{Additional remarks}

Since the appearance of the first version of this paper, objections 
have been raised that the conclusions obtained above are not correct 
and are in conflict with the results previously obtained 
in literature. More precisely, it has been claimed that there is 
actually no conceptional difference between the case of the spin-two 
field and the spin-one field (Maxwell theory), and in particular that 
in both theories, the system is completely specified by the propagating 
fields and the constraints alone. Therefore, we will try, in this section, 
to explain in more detail and from a slightly different 
 point of view why we feel 
that there are fundamental differences, and in particular to clarify 
the meaning of our conclusion that there is no Coulomb  gauge for 
the spin-two field. 

The Hamiltonian {\it reduction} of the spin-two theory has been 
carried out in \cite{deser}, and generalized to General Relativity in
 \cite{adm1}. The word reduction has been put into italics to underline 
the difference to the Dirac approach 
(as used in \cite{baaklini} and \cite{dewitt}, respectively), 
where the constraints are imposed on the states.  

We will confine ourselves mainly to the linear spin-two theory. The results 
of \cite{adm1} supply very strong support for our conclusion, that the 
situation  in the self-interacting case is fundamentally the same. Indeed, 
the authors find that the theory contains two propagating degrees of 
freedom, which according to our counting argument of the previous section, 
means that there should remain one additional equation that is not part of 
the constraints. The strong similarity between General Relativity and 
its linear counterpart with respect to the choice of the 
canonical (propagating) variables has been very explicitely underlined in 
section 4.4.~of the classical ADM article \cite{adm2}: 
\begin{quote}
We will now see the usefulness  of the linearized theory in suggesting 
the choice of canonical variables for the full theory. Since the 
identification is made from the bilinear part of the Lagrangian
 $\pi^{ij}\partial_t g_{ij}$, 
which is the same as for the linearized  theory, the greater complexity 
of the full theory, i.e., its self-interaction, is to be found only 
in the non-linearity of the constraint equations.
\end{quote}
Let us  therefore  turn to the spin-two field. First, it should be noted that 
the reduction in \cite{deser} is based on a first order (Palatini type) 
formulation of the theory. This does not affect the issues we discuss here, 
 since, after all, the Hamiltonian approach is in fact a first 
order reformulation of the Lagrangian theory anyway. Greater ease 
in the reduction process has been obtained in \cite{deser} by the use 
of the decomposition of symmetric tensors 
$A_{\mu\nu} = {}^{\text{\tiny TT}}\!A_{\mu\nu} 
+{}^{\text{\tiny T}}\!A_{\mu\nu} + a_{\mu,\nu} + a_{\nu,\mu}$, where the 
first term is traceless-transverse and the second one transverse. The authors 
then solve the second constraint in (\ref{10}), $h^{\mu\nu}_{\ \ ,\nu,\mu} - 
\Delta h= 0$, with 
 $h_{\mu\nu} = \HH_{\mu\nu} + \xi_{\mu,\nu} + \xi_{\nu,\mu}$. 
This is consistent with our results, namely the solution (\ref{11}), 
if we insert the expression for $\tilde h_{\mu\nu}$ in terms 
of $\HH_{\mu\nu}$ that can be obtained from (\ref{14b}) solving for the 
first term at the r.h.s..  This form is 
particularly useful, since it shows that $h_{\mu\nu}$ is actually 
traceless-transverse, up to a gauge transformation. 

Let us therefore adopt the gauge where $h_{\mu\nu} = \HH_{\mu\nu}$. 
There are not further transformations (\ref{24}) possible that 
involve $\xi^{\mu}$. Let us further impose the gauge $h^{0\mu}_{\ \ ,\mu} = 
0$, which exhausts the last gauge freedom for $\xi^0$. Let us now 
consider the field equations (\ref{21})-(\ref{23}). The first is now 
identically satisfied, since we have already solved the constraint. 
The second reduces to $\Delta h_{0\mu} = 0$, i.e., $h_{0\mu} = 0$. 
The traceless-transverse part of (\ref{23}) remains unchanged and 
leads to the wave equation for the propagating fields. At this point, 
we have completely fixed the gauge, we have solved the constraint 
equations (\ref{21}) and (\ref{22}), and we have also obtained the 
dynamical equation for the radiative field. Thus, if it is true that 
there is no difference to the case of Maxwell's theory, then the 
system should now be completely determined. In other words, $h_{00}$ 
remains arbitrary. But this is obviously not in accordance with the 
Lagrangian field equations, since the trace of (\ref{23}) leads to 
the additional equation $\Delta h_{00} = 0$, i.e., $h_{00} = 0$. 
That is why we claim that the theory is not completely described by 
the constraints and the propagating parts alone. It is interesting 
to note that the previously {\it dynamical} equation $\ddot h + \Delta h_{00} 
- 2 \dot h^{0\lambda}_{\ \ ,\lambda}=0$ (see (\ref{27})) 
has been reduced, in the particular 
gauge, to a {\it constraint} equation, $\Delta h_{00} = 0$. This explains 
why this equation is neither a true constraint, nor a true dynamical equation, 
because, depending on the gauge one adopts, it appears in one or the other 
form. For the same reason, it does not turn up in the Faddeev-Jackiw 
reduction.

What does this have to do with the Coulomb gauge? With the above 
gauge choice, we have $\Delta h_{0\mu} = \Delta h_{00} = 0$, and 
$h_{\mu\nu} = \HH_{\mu\nu}$, which looks like a perfect Coulomb gauge. 
It is not a Coulomb gauge however, simply because it is not gauge, 
but it is a combination of a gauge and a solution. In fact, it is 
the direct analogue of the radiation gauge $A_i = (0, {}^{\text{\tiny T}}\!
A_{\mu})$ in Maxwell's theory, which is a combination of the 
transverse gauge $A_{\mu}= {}^{\text{\tiny T}}\!A_{\mu}$ and the solution 
$A_{0} = 0$ of the constraint equation $\Delta A_0 = 0$. So, what we 
have actually established above is the existence of the radiation gauge 
$h_{00}  = h_{0\mu} = 0$ and $h_{\mu\nu} = \HH_{\mu\nu}$. But this 
is not really exciting though, it has been know at least since the 
  works on weak gravitational 
waves by  Einstein and Rosen and is also shown in almost any textbook 
on General Relativity. This has nothing to do with the Coulomb gauge. 

The difference between  Maxwell's theory 
and  the spin-two field lies in the following. 
In the Maxwell case, we can solve the constraint equation $F^{\mu0}_{\ \ ,\mu} 
= 0$ with  $A_0 = (1/\Delta) \dot A^{\mu}_{\ ,\mu}$. 
This is completely independent 
of the gauge choice. In particular, we can  impose  the transversality  
condition on  $A^{\mu}$. But obviously, the transversality condition can 
be imposed independently of whether $A_{\mu}$ is a solution of the constraint 
or not. Why is that of importance? Well, if we have a source, 
then the solution of the constraint reads 
$A_0 = (1/\Delta) (\dot A^{\mu}_{\ ,\mu} + \rho_{el})$, but just as before,  
we can impose the gauge $A^{\mu}_{\ ,\mu} = 0$, meaning that apart from 
$A_0$ determined as above by the constraint, 
we have only to deal with the tranverse 
part of $A_{\mu}$, which is determined by the dynamical equations.

On the other hand, in the spin-two theory, the constraint reads 
$h^{\mu\nu}_{\ \ ,\mu,\nu} - \Delta h = 0$. As we have seen, this can 
be solved as $h_{\mu\nu} = \HH_{\mu\nu} + \xi_{\mu,\nu} + \xi_{\nu,\mu}$, 
and it is possible to impose the traceless-transverse condition. 
But now, this condition can only be imposed on the solutions of 
the constraint, and not independently of the constraint, as in 
Maxwell theory. In general, we have $h^{\mu\nu} = \HH_{\mu\nu} + 
{}^{\text{\tiny T}}\! h_{\mu\nu} + \xi_{\mu,\nu} + \xi_{\nu,\mu}$, which 
does not allow to choose a traceless-transverse gauge.  

The point is that, as soon as we have source terms, the constraint 
will change, and so will the solution to the constraint. It is not 
hard  to see that, if we have a source $T_{00} = \rho$ in  
the field equation (\ref{21}), i.e., in the constraint equation, 
then the solution to the constraint can be brought into the form 
 $h_{\mu\nu} = 
\tilde h_{\mu\nu} - \frac{1}{2} \eta_{\mu\nu} (\nabla_{\alpha} \nabla_{\beta}/
\Delta) \tilde h^{\alpha\beta} + \eta_{\mu\nu}\ 
\rho/\Delta$, or, inserting the expression 
for $h_{\mu\nu}$ in terms of $\HH_{\mu\nu}$ with the help of (\ref{14b}), 
$h_{\mu\nu} = \HH_{\mu\nu} + \xi_{\mu,\nu} + \xi_{\nu,\mu} 
+ \eta_{\mu\nu}\ \rho/\Delta$, where $\xi_{\mu}$ is given in terms of 
$\tilde h_{\mu\nu}$. 
But this is not equal to $\HH_{\mu\nu}$ up to  a gauge transformation. 
Thus, we cannot reduce $h_{\mu\nu}$ to the propagating components only. 

To summarize, the traceless-transverse condition cannot be
imposed independently  of the solutions to the constraint. 
In other words, there is no 
traceless-transverse gauge in the strict sense.  
There are only solutions (namely  the purely radiative solutions) 
that can be brought into that form. This is the reason why in textbooks 
on General Relativity, explicit reference to wave solutions has to be made 
in order to show the traceless-transverse nature of gravitational waves.  
(Quite in contrast to the Maxwell case, where the transversality condition 
can be trivially imposed.) 

From a more mathematical standpoint, we can recognize the difference 
to Maxwell's theory in the following. In the Hamiltonian (\ref{8}), 
although the terms in $h_{\mu\nu}$ are gauge invariant, the 
terms $\pi^{\mu\nu}\pi_{\mu\nu} - \frac{1}{2} \pi^2$ are not, as
 is easily shown from $\delta \pi_{\mu\nu} = - \xi^0_{\ ,\mu,\nu} 
+ \eta_{\mu\nu} \Delta \xi^0$ under (\ref{24}). Otherwise stated, 
if we use the decomposition $\pi_{\mu\nu} = \PP_{\mu\nu} + \P_{\mu\nu} 
+ p_{\mu,\nu} + p_{\nu,\mu}$,  the Hamiltonian will  not be independent 
of the gauge field $\P_{\mu\nu}$. Rather,  we find $\delta(\pi_{\mu\nu} 
\pi^{\mu\nu} - \frac{1}{2} p^2) = - 2 \xi^0 \pi^{\mu\nu}_{\ \ ,\mu,\nu} 
$, which is proportional to the constraint. Thus, the Hamiltonian 
is invariant only on the constraint surface, quite in contrast to 
 the Maxwell case, where the gauge fields, i.e., the longitudinal modes 
of $A_{\mu}$ do not appear in the Hamiltonian at all. Obviously, the 
above variation is compensated by the variation of $h_{0\mu}$ in (\ref{8}). 
The fact that both terms are not independently gauge invariant, however, 
makes clear that we cannot require the gauge fields not to appear 
in the Hamiltonian and to have $h_{0\mu}$ and $h_{00}$ as arbitrary 
Lagrange multipliers at the same time, as seems to be the viewpoint in 
\cite{deser}.  It is either one or the other. 
For the exact same reason, those fields are  interrelated by 
the {\it missing} equation.  

If this is still not convincing, here is an explicit example 
that shows that the system of constraints and dynamical (propagating) 
field equations is under-determined. Consider the solution given by  
$h_{\mu\nu}= \pi_{\mu\nu} = 0$, $h_{0\nu} = 0$, 
$h_{00} = \exp[-r]$, with $r = \sqrt{x^2+ y^2+z^2}$.  
The constraints $h^{\mu\nu}_{\ \ ,\mu,\nu} + \Delta h = 0$ and 
$\pi^{\mu\nu}_{\ \ ,\mu}= 0$ are trivially satisfied, and so are the
dynamical equations (\ref{18}). Nevertheless, 
this is not a solution of the field equations 
(\ref{21})-(\ref{23}). The trace of $G_{\mu\nu}$ does not vanish. 
In other words, the {\it missing} equation $\dot \pi = \Delta h_{00}$, 
or, in configuration space, $\ddot h + \Delta h_{00} - 
2 \dot h^{0\mu}_{\ \ ,\mu}= 0$ (see (\ref{27}))   is not satisfied, 
since $\Delta h_{00} \neq 0$.

Those considerations support further our conclusions that there are 
fundamental differences between the spin-one theory and the spin-two 
theory, namely the absence of a Coulomb gauge in the latter, as well 
as the fact that the theory is not completely specified  by the constraints 
and the dynamical (propagating) equations alone. For those reasons, 
the Faddeev-Jackiw procedure does not lead to the complete set of 
independent field equations.

\end{document}